# Looking for Hidden Photon's Cold Dark Matter By Multi-Cathode Counter

**A.V.Kopylov[1], I.V.Orekhov and V.V.Petukhov**

*Institute for Nuclear Research of Russian Academy of Sciences,
117312, Prospect of 60$^{th}$ October Revolution 7A, Russia
E-mail*: beril@inr.ru

ABSTRACT: The dark rate of single-electrons emitted from a cathode of a multi-cathode counter may be interpreted as the effect from Hidden Photons (HPs) of Cold Dark Matter (CDM). As a tentative approach the measurements have been performed at different temperatures of the counter. Preliminary results of measurements by using a counter with a copper cathode are presented. An upper limit of $4.3 \cdot 10^{-23}$ for parameter $\chi^2$ at $2\sigma$ has been obtained for hypothetical conversion of hidden photon to real photon at the metallic surface with the resulting emission of single electron. The work is considered as a pure illustration of the method and the perspectives are discussed for future search of HPs by a multi-cathode counter.

## 1. Introduction

Hidden photons (HPs) were proposed by L. B. Okun [1] in 1982 as a possible modification of electrodynamics. The hypothetical HPs are also interesting as one of the alternatives for Cold Dark Matter (CDM). The fact that mass of a real photon is zero it follows just from Lorentz calibration. For massive hidden photons the expression for Lorentz calibration should be also modified so that mixing between these two states becomes nearly impossible as being prohibited by pure kinematics. But this can be different at the interface of metal-dielectric where the conversion of HPs from the dielectric side into photons at the metallic surface can't be excluded as a pure hypothetical process. In experiment this process could be manifested by the emission of single electrons from the metallic surface. It would be very interesting to test this in experiment. The probability of this conversion could be evaluated as a parameter $\chi^2$ which could be very small if not negligible. To observe this effect in experiment one should use a detector with relatively large surface and high sensitivity to single electrons emitted from metal. We have developed a special Multi-Cathode Counter (MCC) which has relatively large (2400 cm$^2$) surface, high (~10$^5$) gas amplification which enables to register single-electrons with high sensitivity and has low dark count rate of single electrons. Here we present preliminary results of measurements by this detector and the interpretation of these results in terms of the possible effect from HPs.

## 2. Experimental setup and selection of "true" events.

The general view and a design of MCC were presented in [2, 3]. Here they are shown at Figure 1. To reduce the background from the surrounding γ-radiation the counter has been placed in a special cabinet having as a passive shield the steel slabs in total 30 cm from all sides. The measurements performed inside and outside of the steel cabinet have shown that the background of single-electron events from external γ-radiation has been attenuated till the negligible level in comparison with the measured effect.

---

[1] Corresponding author.

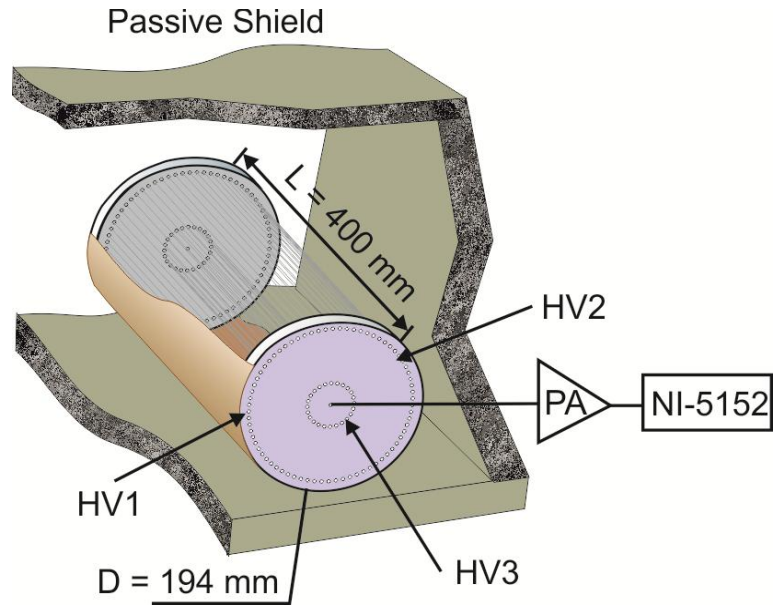

Figure 1. Schematic of MCC.

The counter is mounted in a hermetic stainless steel body with all voltages supplied by means of hermetic feedthrough. It is filled with a mixture of Ar + 10% $CH_4$ at a pressure of about 0.2 MPa. The counter has several cathodes with negative high voltage applied to each cathode. First cathode is made of a copper sheet which is a metallic surface emitting single-electrons upon the absorption of UV light during calibration or as a result of a hypothetical conversion of HPs into usual photons absorbed by a surface of a metal cathode as we are searching for in this experiment. At a distance of 8 mm from the first cathode a second one is arranged made of array of tighten nichrome wires 50 microns in diameter with an interval between wires of 8 mm. Different high voltages are applied to this cathode: in configuration 1 the potential of second cathode do not prevent electrons to move towards central counter while in configuration 2 the potential applied to second cathode acts as a barrier, so that electrons are scattered back towards first cathode. Cathode 3 serves as a cathode of central counter with high ($\sim 10^5$) gas amplification which enables to detect single-electrons. Central wire of the counter is made of gold-plated W-Re alloy 25 microns in diameter. The signal from central wire is fed to the input of charge-sensitive preamplifier and then to the input of 8-bit digitizer. All data collected during measurements are stored on a disk and are analyzed off-line. The effect is evaluated by the difference of the count rates in first and second configuration. It was presumed that in first configuration we measure the count rate $R_1$ of single-electrons emitted: (1) from the fist cathode plus (2) from the surfaces at the ends of the counter plus (3) from the multiple wires inside the counter while in the second configuration we measure the rate of single-electrons only from second and third components. Electrons emitted from first cathode are scattered back in second configuration and do not contribute to the total count rate $R_2$. So difference $\Delta R = R_1 - R_2$ of the count rates in first and second configurations should give the net effect from single-electrons emitted from an external cathode. To find the efficiency of the counter for single-electron measurements the calibration has been performed by UV mercury vapor lamp through a quartz window in the side of the counter. The data treatment was performed in off-line. For the selection of "true" events a selection has been performed in space of three parameters: amplitude of the pulse, the duration of the leading edge of the pulse and a parameter β which describes prehistory of the event and is proportional to a first derivative of a baseline, approximated by a straight line during 50 μsec before leading edge of the pulse. The efficiency was estimated as the probability for the pulse to belong to ROI box of this 3-parameter space. It was found to be (88 ± 6) %. To reduce the influence of the noise on counting only intervals with a baseline deviation from zero not more than 5 mV were taken into account with a proper correction fort a live time of counting which was found to be about 54%.

## 3. Loss of electrons due to attachment.

The number of electrons lost during diffusion of electrons from the external cathode of the counter to central counter was estimated following mechanism of attachment proposed by Bloch and Bradbury and developed by Herzenberg (BBH model) [4, 5] for electrons of small energy. The number of free electrons in a gas which contains electronegative impurities (in our case this is an oxygen impurity) is decreasing exponentially:

$$N(t) = N(0) \cdot e^{-At} \quad (1)$$

here N(t) – a number of electron at time t, N(0) – the initial number of electrons, A - "the velocity of attachment". It can be described by the expression

$$A = P(M) \cdot P(O_2) \cdot C_{O2,M} \quad (2)$$

here P(M), P($O_2$) – the pressure of working gas and the one of oxygen, $C_{O2,M}$ – the coefficient of attachment, which does not depend upon the pressure of gas and of impurity in BBH model.
We have done the calculation of electric field strength in our detector by means of Maxwell16 (Figure 2).

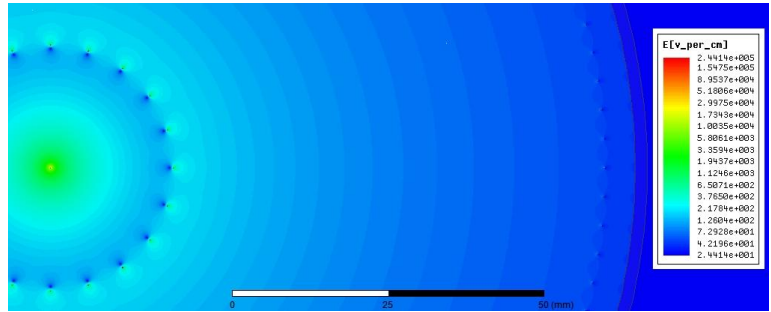

Figure 2. The calculated electric field strength across the counter.

Using the data on dependence of the drift velocity of electrons upon E/p in argon-methane mixture [6] and on coefficient of attachment $C_{O2,M}$ from [7] following the expressions (1) and (2) we determined that probability for electron to be attached while drifting on the way from external cathode to a central counter was about 1%, i.e. it can be neglected.

## 4. Possible effect from HPs.

A detector is searching for single-electrons emitted from the surface of metallic sheet upon the hypothetical conversion of HPs into real photons absorbed by the surface of a metal. If DM is totally composed of hidden photons, the power collected by antenna [8] (in our case – by a cathode of the counter)

$$P = \chi^2 \rho_{CDM} A_{cath} \quad (3)$$

where: $\rho_{CDM} \approx 0.3$ GeV/cm³ is the energy density of CDM which is taken here to be equal to the energy density of HPs and $A_{cath}$ – the cathode's surface of the counter. If all the power is collected by traps at the surface of a metallic cathode and then is emitted by single-electrons,

$$P = \varphi_W \cdot \Delta R \quad (4)$$

where: $\varphi_W$ is the work function of a metal, cathode of the counter is made of, $\Delta R$ is the rate of single-electrons emitted from the cathode. From the expressions (3) - (4) it follows:

$$\chi^2 = \frac{\varphi_W \cdot \Delta R}{\rho_{CDM} \cdot A_{cath}} \quad (5)$$

From here one obtains:

$$\chi^2 = 3.3 \cdot 10^{-24} \left(\frac{\varphi_W}{eV}\right)\left(\frac{\Delta R}{Hz}\right)\left(\frac{\rho_{CDM}}{0.3\ GeV/cm^3}\right)^{-1}\left(\frac{A_{cath}}{m^2}\right)^{-1} \qquad (6)$$

Here one should note that the expression (5) is obtained in the supposition that all single-electrons emitted by metal are the result of conversion of HPs into real photons absorbed by a surface of a metal what is not fulfilled here. Single-electrons can be also emitted as a result of several other sources. The defects on the surface of wires, protrusions and spots of heterogeneity on the surface of a cathode sheet etc may generate single-electrons. Taking this into consideration one should accept $\chi^2$ found from (5) only as an upper limit. To improve the limit one should modify the experiment in a way to diminish the effect from extraneous sources. One of the ways would be also to lower the temperature of the detector. However, there are some indications [9-12], that some detectors exhibit unusual temperature dependence: the dark rate of single-electrons does not get decreased but contrary to the expectations gets increased with lowering the temperature. So far this did not find any satisfactory explanation. One of possible explanation could be that some part of the energy absorbed as a result of conversion of HPs at the surface of a metal cathode is eradiated by thermal radiation and this process is attenuated by lowering the temperature of a detector. If this is true it means that to get a right result one should make measurements at cryogenic temperature as low as 4 K. It is very problematic for MCC which is a gaseous detector. But to measure the temperature dependence of the count rate of single-electrons by means of MCC at somewhat higher temperatures may become very indicative for what should be observed at lower temperatures. So the result obtained from expression (6) has to be taken very cautiously with all limitations discussed here. Apparently these questions need a further study. It would be also very helpful to measure the effect for different metals.

## 5. The results obtained.

The measurements have been performed during 78 days at 26°C, 31°C and 36°C. During this time 15 TB of information has been obtained on the shapes of pulses of different amplitudes. Then the pulses were selected by amplitude, duration of a leading edge and prehistory of the event. Figure 3 shows the region of interest (ROI) used for selection of "true" pulses and distribution of the events for all three sets of measurements. The selection of ROI has been performed from data obtained by calibration of the counter by means of UV Mercury vapor lamp. As the "true" pulses were accepted only the pulses having amplitudes from 3 till 30 mV, duration of the front edge from 2 till 25 μs and a parameter β which is dependent from prehistory of the event from -0.1 till 0.1 which excludes events with strong distortion of the baseline prior the front edge of the pulse. One can see that all three sets of measurements have similar distribution of pulses which proves the stability of the measurements.

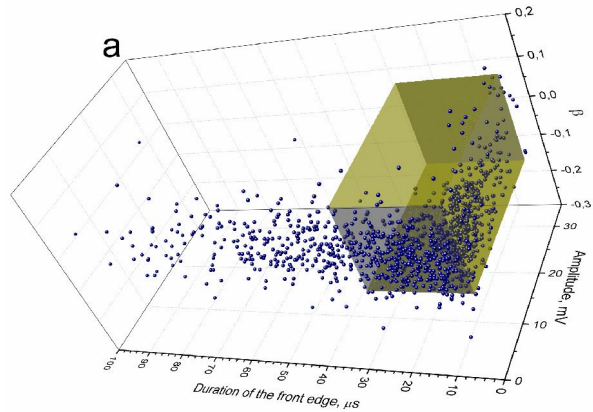

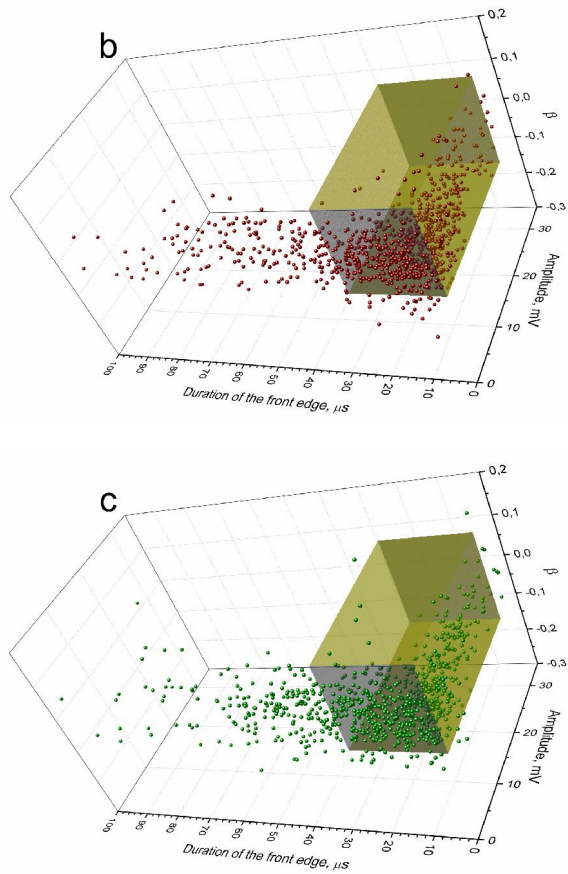

Figure 3. The distribution of the events in 3D space of parameters: amplitude of the pulse, duration of the front edge of the pulse and parameter β which is dependent upon the prehistory of the event. The distributions are presented for measurements at different temperatures: 26°C(a), 31°C(b), 36°C(c).

Figure 4 shows the results obtained by these temperatures. The efficiency of counting as a probability of the event to belong to ROI was estimated from calibration to be (88 ± 6) %.

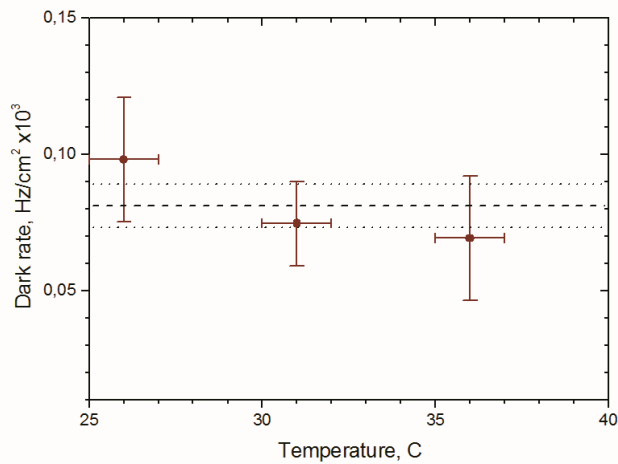

Figure 4. The distribution of events in sets 26°C, 31°C, 36°C. Dashed line – the average value obtained for all measurements, by point line are depicted levels ±1σ from average value.

One can see that all three sets have close distributions. After selection of "true" pulses we obtained for $r_{MCC} = \Delta R/A_{cath}$: $(0.98 \pm 0.22) \cdot 10^{-4}$ Hz/cm$^2$, $(0.75 \pm 0.15) \cdot 10^{-4}$ Hz/cm$^2$ and $(0.69 \pm 0.23) \cdot 10^{-4}$ Hz/cm$^2$ for these temperatures correspondingly. The fact that the count rate was not revealing a clear increase with temperature can be taken as evidence that there was negligible contribution of thermal emission. As an average value it was obtained $r_{MCC} = (0.81 \pm 0.08) \cdot 10^{-4}$ Hz/cm$^2$. From here following the expression (6) an upper limit was obtained at $2\sigma$: $\chi^2 < 4.3 \times 10^{-23}$. This result is obtained by very tentative approach and it can be taken only as a demonstration of the possibility of the method. The task for future experiment is to increase the accuracy of measurements and to measure precisely the temperature dependence of the effect.

## 6. Discussion.

The limit for $\chi^2$ obtained by means of this technique depends critically upon the dark rate of the detector used for counting of single electrons. There are ways to improve detector by using new materials, new technologies of treatment of metallic surface etc. The experiment can be improved also by lowering the temperature of the detector to reduce the effect from thermal noise. Moving along this line would enable to improve this result. By the moment a new counter of a developed design has been fabricated with a cathode made of aluminum alloy and now it is tested prior to start measurements. For near future we are planning also to construct MCC using materials having relatively high work function like nickel or platinum.

**Conflict of Interests**

The author declares that there is no conflict of interests regarding the publication of this paper.

**Funding Statement**

The work was supported by Federal Agency of Scientific Organizations, Russia.

**Acknowledgement**

The authors are grateful to E.P.Petrov for active participation in the construction of the counter.